\documentclass[twocolumn,english]{revtex4}
\usepackage{times}
\usepackage[T1]{fontenc}
\usepackage[latin1]{inputenc}
\usepackage{prettyref}
\usepackage{amsmath}
\usepackage{graphicx}
\usepackage{amssymb}

\makeatletter
\usepackage{babel}
\makeatother
\begin{document}

\title{Dynamics of Epidemics}

\author{M Marder}

\address{Center for Nonlinear Dynamics and Department of Physics\\
The University of Texas at Austin, 78712,USA}

\email{marder@mail.utexas.edu}

\pacs{89.75.Da,89.75.Hc,64.60.Ak}

\begin{abstract}
This article examines how diseases on random networks spread in time.
The disease is described by a probability distribution function for
the number of infected and recovered individuals, and the probability
distribution is described by a generating function. The time development
of the disease is obtained by iterating the generating function. In
cases where the disease can expand to an epidemic, the probability
distribution function is the sum of two parts; one which is static
at long times, and another whose mean grows exponentially. The time
development of the mean number of infected individuals is obtained
analytically. When epidemics occur, the probability distributions
are very broad, and the uncertainty in the number of infected individuals
at any given time is typically larger than the mean number of infected
individuals.
\end{abstract}
\maketitle

\paragraph*{Introduction}

A series of papers by Strogatz, Watts\cite{Watts1998}, Vespignani\cite{Pastor-Satorras2002,Pastor-Satorras2001},
Meyers\cite{Meyers2003,Meyers2005,Meyers2006}, Newman\cite{Newman2003b,Newman2002d,Newman2001,Moore2000,Newman1999},
Stanley\cite{Liljeros2001}, Barab\'asi\cite{Albert2002} and collaborators
applies methods from graph theory and percolation theory to the spread
of disease on random networks. These papers mainly study the final
state of a population once the disease has run its course, with all
individuals susceptible but uninfected, or recovered. Here I show
how to apply the same analytical techniques to dynamics of the epidemic
and find how the number of infected individuals varies in time. 

A starting point for this study was to clear up a technical point
arising when an epidemic is possible, but not certain. Newman, Strogatz
and Watts\cite{Newman2001} find a probability distribution function
$P_{k}$ that $k$ individuals have been infected. and they show that
$u\equiv\sum_{k}P_{k}<1.$ They determine $u$ from a self-consistent
equation, and interpret this distribution function as describing the
probability of a finite outbreak that does not grow to system size.
The remaining probability is contained in an outbreak that fills the
whole system. This interpretation is puzzling. Since $k$ can have
any size, why does $P_{k}$ describe only finite outbreaks? How does
the self--consistent equations determining $u$ figure out how to
find only these finite outbreaks, and discard the larger ones? The
authors assert that the system-size outbreaks would contain loops
that invalidate the formalism they are employing, but how does the
formalism know this? These questions are resolved when one examines
the probability distribution after $n$ times steps, $P_{k}^{(n)}.$
One finds that the probability distribution is the sum of two pieces.
The first piece $Q_{k}^{(n)}$ converges to a time-independent function
$Q_{k}$ in the long-time limit, with $\sum_{k}Q_{k}<1$. The second
piece $R_{k}^{(n)}$ never stops evolving. Its mean and width grow
exponentially. So long as the mean of $R_{k}^{(n)}$ is much smaller
than the total system size, it can be described by standard generating
function techniques, and this description is not invalidated by the
presence of loops. Thus, the generating function formalism has been
finding $Q_{k}$ and the reason this function emerges is that $P_{k}^{(n)}$
converges to $Q_{k}$ pointwise, although at any given time step $n$
a finite fraction of $P_{k}^{(n)}$ is contained in a very broad tail
of the distribution that has formed out in front of $Q_{k}$. Techniques
essentially identical to those used previously to describe $Q_{k}$
can be used to analyze $R_{k}^{(n)}.$ In particular, one can find
closed-form expressions for the mean number of people infected at
time $n.$ When an epidemic is possible, both the mean and width of
$R_{k}^{(n)}$ grow exponentially in time. In general, ones uncertainty
about precisely how many people will be infected in the future grows
as fast as or faster than the number of diseased individuals.

\paragraph*{Dynamical Equations}

Consider a random network in which the probability distribution of
nodes with $k$ edges is $p_{k}.$ Following Newman, Strogatz, and
Watts\cite{Newman2001}, the generating function for the distribution
of nodes is\begin{equation}
G_{0}(x)\equiv\sum_{k=0}^{\infty}p_{k}x^{k}.\label{eq:DE0}\end{equation}
Consider choosing a random edge in the system. The probability that
the node reached by this edge will have $k$ new edges in addition
to the one chosen to start with is generated by\begin{equation}
G_{1}(x)\equiv\frac{G_{0}'(x)}{G_{0}'(1)}.\label{eq:DE1}\end{equation}
Consider conventional Susceptible-Infected-Recovered dynamics on this
network\cite{Newman2001}. At each time step, uninfected nodes connected
by an edge to infected nodes become infected in turn. Let $P_{k}^{(n)}$
give the probability that a grand total of $k$ individuals has been
infected after $n$ time steps, and let the generating function for
$P_{k}^{(n)}$ be\begin{equation}
H^{(n)}(x)\equiv\sum_{k=0}^{\infty}P_{k}^{(n)}x^{k}.\label{eq:DE2}\end{equation}
 Imagine starting with a single infected individual. At step $0$,
one has $H^{(0)}=x.$ At the next time step, the generating function
for the total number of individuals infected is \begin{equation}
H^{(1)}(x)=xG_{0}(x),\end{equation}
since $G_{0}(x)$ gives the probability that a given node has 0, 1,
2, $\dots$ edges, and one multiplies by $x$ because one began with
one infected individual. Each of the edges departing the first one
reaches some other node. The probability it will have $k$ additional
edges leaving it is given by $G_{1}(x).$ Using the \emph{powers}
property in Section IIA of Ref. \cite{Newman2001}, one has\begin{equation}
H^{(2)}(x)=xG_{0}(xG_{1}(x)).\end{equation}
Continuing in this fashion, one has\begin{equation}
H^{(n)}(x)=H^{(n-1)}(xG_{1}(x)).\end{equation}
This expression is inconvenient form for numerical work, so define
instead\begin{subequations}\begin{eqnarray}
F^{(0)}(x) & = & 1\label{eq:DE5;a}\\
F^{(n)}(x) & = & G_{1}(xF^{(n-1)}(x))\label{eq:DE5;b}\\
H^{(n)}(x) & = & xG_{0}(xF^{(n-1)}(x))\label{eq:DE5;c}\end{eqnarray}
\label{DE5}\end{subequations}To extract the probability distribution
function from a generating function $H(z)$, note that from Cauchy's
theorem\begin{equation}
P_{k}=\frac{1}{2\pi i}\oint\frac{dz}{z^{k+1}}H(z)=\int_{0}^{1}d\theta\: e^{-2\pi ik\theta}H(e^{2\pi i\theta}).\end{equation}
Suppose now that $H$ has been evaluated around the unit circle at
$M$ points, with $\theta_{l}=l/M,\ l\in[0,M-1]$, and let\begin{equation}
H_{l}=H(e^{2\pi i\theta_{l}}).\end{equation}
Then one has\begin{equation}
P_{k}=\frac{1}{M}\sum_{l=0}^{M-1}e^{-2\pi ikl/M}H_{l}=\frac{1}{M}\mbox{DFT}(H,-1)[k]\label{eq:DE7}\end{equation}
 where the last expression means that one takes the $k$'th element
of the inverse discrete Fast Fourier Transform. Using Eq. \prettyref{eq:DE5;c}
and employing Eq. \prettyref{eq:DE7} to obtain probabilities $P_{k}$,
one easily obtains hundreds of iterates of the map, for hundreds of
thousands of values of $k.$ 

\begin{figure}[h]
\includegraphics[width=0.9\columnwidth]{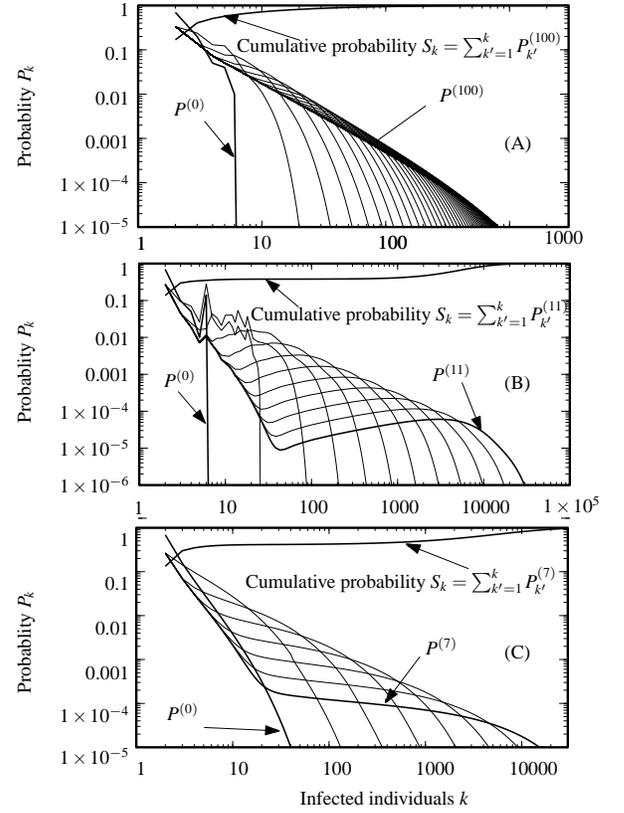}

\caption{(A)Dynamical evolution of Eq.\prettyref{eq:DE5;c} in a case where
the average number of second neighbors $z_{2}$ is less than the average
number of neighbors $z_{1}$ and there is no epidemic. The map is
iterated 100 times.(B) Dynamical evolution of Eq.\prettyref{eq:DE5;c}
in case where $z_{2}>z_{1},$ so one expects the existence of a giant
component. The map is iterated 12 times. (C) Similar to (B), but now
using a broader probability distribution. The epidemic grows quickly
and only 7 iterations are displayed.\label{cap:Dynamical-evolution-of}}
\end{figure}

\begin{figure}
\includegraphics[width=0.9\columnwidth]{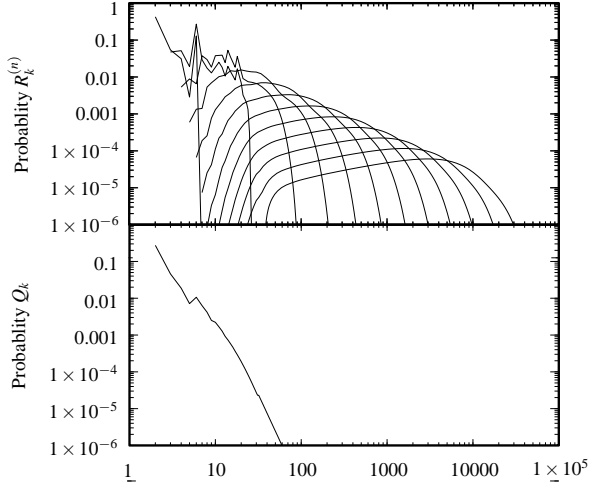}

\caption{Decomposition of the data in Figure \ref{cap:Dynamical-evolution-of}
(B) into static and growing components $Q_{k}$ and $R_{k}^{(n)}.$
This is done by computing $P_{k}^{(11)}$, setting $Q_{k}=P_{k}^{(11)}$
for $k\leq32,$ fitting $Q$ to a power law for for $k>32,$ and subtracting
$Q_{k}$ obtained this way from distributions $P_{k}^{(1)}\dots P_{k}^{(11)}$.
\label{cap:Decomposition-of-the}}
\end{figure}

\paragraph*{Static and growing distributions}

Some results of solving Eqs. \prettyref{eq:DE5;c} appear in Figure
\ref{cap:Dynamical-evolution-of}. Figure \prettyref{cap:Dynamical-evolution-of}
(A) shows distributions resulting from the polynomial $G_{0}(x)=.7x+.2x^{2}+.05x^{3}+.04x^{4}+.01x^{5}$.
The threshhold for an epidemic is determined by $z_{2}>z_{1}$, \cite{Molloy1995,Molloy1998,Newman2001,Pastor-Satorras2002}
where $z_{1}=G_{0}'(1)$ is the average number of neighbors of each
node, and $z_{2}=G'_{1}(1)z_{1}$ is the average number of second
neighbors. In the present case, $z_{1}=1.46$ and $z_{2}=1.38$, so
the infection is contained, and the probability distribution converges
to a definite limit enclosing unit probability. The upper curve shows
the cumulative sum $S_{k}=\sum_{k'=1}^{k}P_{k'}^{(100)}$. The mean
number of people infected after 100 iterations is 27, but the distribution
is broad; for example, there is a 1\% chance that more than 480 people
will be infected. Figure \ref{cap:Dynamical-evolution-of} (B) shows
results from the polynomial $G_{0}(x)=.7x+.1x^{2}+.05x^{3}+.01x^{4}+.14x^{5}$,
which gives $z_{1}=1.79$, $z_{2}=3.42$. Since $z_{2}>z_{1},$ an
epidemic is possible. One can compute the probability of an epidemic
spiraling out of control  following \cite{Newman2001}; also see Eq.
\prettyref{eq:ulimit}. There is a root of $G_{1}(u)-u$ at $u=.492$
and $G_{0}(u)=0.3790$. This computation predicts a 37.9\% chance
that the disease will run its course without becoming an epidemic.
The upper curve in Figure \ref{cap:Dynamical-evolution-of} (B) shows
the cumulative sum $S_{k}=\sum_{k'=1}^{k}P_{k'}^{(11)}$, and there
is a broad plateau where this sum has reached .38. The mean number
of infected individuals after 11 iterations is 4650 but there is a
1\% chance more than 26000 people will be infected. Figure \ref{cap:Dynamical-evolution-of}
(C) uses the probability distribution $p_{0}=0,$ $p_{k}\propto k^{-\alpha}e^{-k/\kappa}$
with $\alpha=2$ and $\kappa=20.$ Now $z_{1}=1.8,$ $z_{2}=5.3,$
and the epidemic grows even more rapidly. There is a 41\% chance that
the epidemic will be contained. The mean number of infected individuals
after 7 steps is 5500, but there is a 1\% chance that more than 50,000
will be infected. 

Thus when there is the possibility of an epidemic, the probability
distribution does indeed split into two components. The first component
$Q_{k}$ is static in the long-time limit and describes the probability
that spread of disease terminates with a number of infected individuals
much smaller than the total population. The second component $R_{k}^{(n)}$
continues to evolve forever. From a formal point of view, the definition
of $Q_{k}$ is\begin{equation}
Q_{k}=\lim_{n\rightarrow\infty}\int d\theta\: e^{-2\pi ik\theta}H^{(n)}(e^{2\pi i\theta}).\label{eq:Qdef}\end{equation}
For any fixed $k,$ this limit converges. Then $R_{k}^{(n)}$ can
be defined as $R_{k}^{(n)}=P_{k}^{(n)}-Q_{k}.$ One can similarly
decompose the probability distribution resulting from $F^{(n)}$ into
static and evolving components. To see now how the probability of
not participating in the epidemic emerges from self-consistent equations,
define $F^{\infty}(x)\equiv\lim_{n\rightarrow\infty}F^{(n)}(x)$ .
This limit exists for any $x<1,$ since large powers of $x<1$ in
the power series for $F^{(n)}$ suppress the parts of $F^{(n)}$ that
are continuing to evolve. Return to \prettyref{eq:DE5;b}, and write
\begin{align}
\lim_{x\rightarrow1}\lim_{n\rightarrow\infty}F^{(n)}(x)-G_{1}(xF^{(n-1)}) & =0\nonumber \\
\Rightarrow\lim_{x\rightarrow1}F^{\infty}(x)-G_{1}(xF^{\infty})=0\nonumber \\
\Rightarrow u=G_{1}(u)\quad\mbox{with\quad}u\equiv\lim_{x\rightarrow1}F^{\infty}(x).\label{eq:ulimit}\end{align}
Finally $G_{0}(u)=\lim_{x\rightarrow1}\lim_{n\rightarrow\infty}H^{(n)}(x)$
gives the probability that the disease does not spiral into an epidemic. 

Figure \prettyref{cap:Decomposition-of-the} shows an explicit decomposition
of the data in Figure \prettyref{cap:Dynamical-evolution-of} (B)
into components $Q$ and $R$. The task is carried out by taking the
final curve in \prettyref{cap:Dynamical-evolution-of} (B) and noticing
that it has converged to a static value up to around $k=32$ (the
precise cut point does not matter much) and is continuing to evolve
for larger $k.$ For $k>32,$ $Q_{k}$ is estimated by a power-law
fit. The area under $Q_{k}$ found this way is .3791 which compares
well with the predicted value of .3790.

\paragraph*{Size of infected cluster}

One can work out analytically the average size of the infected/recovered
cluster as a function of time. Note that $F^{(n)}(1)=1$and let\[
M_{n}=\frac{d}{dx}F^{(n)}(x)\vert_{x=1}.\]
Then\begin{equation}
M_{n}=G_{1}'(1)[F^{(n-1)}(1)+M_{n-1}]=\frac{z_{2}}{z_{1}}(1+M_{n-1}).\end{equation}
Using $M_{0}=0$, one can solve this iterated map exactly as a power
series, which has the compact final expression \begin{equation}
M_{n}=\frac{z_{2}}{z_{1}}\sum_{l=0}^{n-1}\left(\frac{z_{2}}{z_{1}}\right)^{l}=\frac{z_{2}}{z_{1}}\left[\frac{1-(z_{2}/z_{1})^{n}}{1-z_{2}/z_{1}}\right].\label{eq:Mn}\end{equation}
Then the average number of individuals in the cluster is\begin{equation}
\langle k\rangle_{n}=\frac{d}{dx}H^{(n)}(x)\vert_{x=1}=1+z_{1}(1+\frac{z_{2}}{z_{1}}\left[\frac{1-(z_{2}/z_{1})^{n-1}}{1-z_{2}/z_{1}}\right].).\label{eq:average}\end{equation}
If $z_{2}<z_{1},$ one obtains the expected result \cite{Molloy1995,Molloy1998,Newman2002d}for
large $n$ that \begin{equation}
\langle k\rangle=1+z_{1}(1+\frac{z_{2}}{z_{1}-z_{2}})=1+\frac{z_{1}^{2}}{z_{1}-z_{2}}.\end{equation}
In the opposite case, $z_{2}>z_{1},$ Eq. \prettyref{eq:Mn} becomes
$M_{n}\approx(z_{2}/z_{1})^{n+1}/(1-z_{2}/z_{1})$ and for large $n$
the average size of the infected population is\begin{equation}
\langle k\rangle_{n}\sim\frac{z_{1}(z_{2}/z_{1})^{n}}{z_{2}/z_{1}-1}.\end{equation}
The width of the distribution is proportional to the mean. The dominant
contribution to $\langle k^{2}\rangle_{n}$ at large $n$ is

\begin{equation}
\sqrt{\langle k^{2}\rangle_{n}-\langle k\rangle_{n}^{2}}\sim\frac{(z_{2}/z_{1})^{n}}{z_{2}/z_{1}-1}\sqrt{z_{2}-z_{1}^{2}+\frac{z_{2}G''_{1}(1)}{(z_{2}/z_{1}-1)}}\end{equation}

\paragraph*{When infection is not certain across an edge}

Newman\cite{Newman2002d} describes the case where infection is not
certain across an edge connecting two nodes, but occurs with probability
$T.$ In this case, the probability of infecting neighbors starting
with a randomly chosen node is generated by\begin{equation}
G_{0}(1+T(x-1)),\end{equation}
the probability of infecting neighbors starting with a randomly chosen
edge, excluding the incoming edge is generated by\begin{equation}
G_{1}(1+T(x-1)),\end{equation}
and employing these two generating functions, the evolution equations\prettyref{eq:DE5;c}
are unchanged, and \prettyref{eq:average} for the average degree
of infection generalizes to\begin{equation}
\langle k\rangle_{n+1}=\frac{d}{dx}H^{(n+1)}(x)\vert_{x=1}=1+\frac{z_{1}^{2}T-z_{2}T(z_{2}T/z_{1})^{n}}{z_{1}-z_{2}T}.\label{eq:average2}\end{equation}
Essentially $z_{2}$ is replaced by $Tz_{2}.$

\paragraph*{Individuals added at each time step}

Another interesting quantity to track is the probability of adding
individuals of varying degree number $k$ at each time step. This
can be done by adding a subscript to the variable $x$ that tracks
the time step at which an individual has entered the cluster. Doing
so one has

\begin{eqnarray*}
H^{(0)}(x_{0}) & = & x_{0}.\end{eqnarray*}
 \begin{equation}
H^{(1)}(x_{0},x_{1})=x_{0}G_{0}(x_{1}),\end{equation}
\begin{equation}
H^{(2)}(x_{0},x_{1},x_{2})=x_{0}G_{0}(x_{1}G_{1}(x_{2})),\end{equation}
Continuing in this fashion, one has\begin{equation}
H^{(n)}(\vec{x})=H^{(n-1)}(x_{0},x_{1},\dots x_{n-1}G_{1}(x_{n})).\end{equation}
One recovers the results in Eq. \prettyref{eq:DE5;c} by removing
all the indices from the variables $x.$ To focus upon the individuals
added to the cluster at step $n,$ just set all variables $x_{l}$
to 1 except the last. Denote by $s_{k}^{(n)}$ the probability that
$k$ individuals have been added at time step $n$, and let $J(x)$
be the generating function for this probability. Then\[
J^{(1)}(x)=G_{0}(x);\quad J^{(n)}(x)=J^{(n-1)}(G_{1}(x)).\]

\begin{figure}[t]
\begin{centering}\includegraphics[width=0.9\columnwidth]{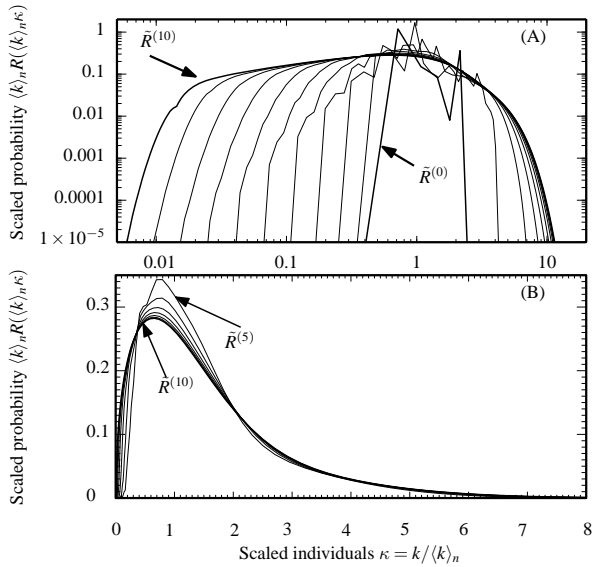}\par\end{centering}

\caption{This plot of $\langle k\rangle P_{\kappa\langle k\rangle}$, using
the generating functions from Figure \ref{cap:Dynamical-evolution-of}
(B) iterated 10 times does appear to be converging on a scaling form.
Convergence for small values of $\kappa=k/\langle k\rangle_{n}$ is
pointwise on a logarithmic scale (A) and uniform on a linear scale
(B). \label{cap:Attempt-to-find}}
\end{figure}

One can now calculate the mean number of people added at each time
step,\[
\delta k_{n}\equiv\sum s_{k}^{(n)}k.\]
\[
\delta k_{1}=z_{1}\]
\[
\delta k_{2}=\frac{z_{2}}{z_{1}}z_{1}\dots\delta k_{n}=\left(\frac{z_{2}}{z_{1}}\right)^{n-1}z_{1}\]

\paragraph*{Scaling form for epidemic}

It would seem natural for the growing part of the probability distribution
$R_{k}$ to adopt a scaling form at long times. To capture the growing
part of the distribution, one computes\begin{equation}
R_{k}^{(n)}\approx\frac{1}{\langle k\rangle_{n}}\tilde{R}(\kappa)\quad\mbox{where}\quad\kappa=k/\langle k\rangle_{n}.\end{equation}
As shown in Figure \ref{cap:Attempt-to-find}, this scaling form does
appear to describe $R$ after sufficiently many iterations, On a log
scale the tail of $\tilde{R}$ for small $\kappa=k/\langle k\rangle_{n}$
converges pointwise, but on a linear scale convergence is uniform.

\begin{acknowledgments}
I would like to thank Lauren Ancel Meyers for introducing me to this
problem, and Shweta Bansal for patiently explaining to me some of
the perplexing features. Thanks also to the National Science Foundation
for support through DMR-0401766.
\end{acknowledgments}
\bibliography{/home/marder/references/epidemic}

\end{document}